\documentclass[a4paper]{spie}  %>>> use this instead for A4 paper
\usepackage[utf8]{inputenc}
\usepackage[]{graphicx}
\usepackage{amssymb,amsmath,psfrag,url}
\usepackage{verbatim}
\usepackage{todonotes}
\usepackage{textcomp}

\usepackage[scientific-notation=false]{siunitx}

\newcommand{\curl}{\nabla \times}
\newcommand{\bbr}{\mathbb{R}}
\DeclareMathOperator*{\argmin}{arg\,min}

\title{Using Gaussian process regression for efficient parameter reconstruction}

\author{
  Philipp-Immanuel~Schneider,\supit{\,a}
  Martin~Hammerschmidt,\supit{\,a} 
  Lin~Zschiedrich,\supit{\,a}
  Sven~Burger\supit{\,ab}
\skiplinehalf
\supit{a}
JCMwave GmbH\\
Bolivarallee~22, 
D\,--\,14\,050 Berlin,
Germany
\smallskip\\
\supit{b}
Zuse Institute Berlin (ZIB)\\
Takustra{\ss}e~7,
D\,--\,14\,195 Berlin,
Germany
\authorinfo{
Corresponding author: P.-I.~Schneider\\
URL: http://www.jcmwave.com
}}
\begin{document}	
\maketitle

\begin{abstract}
Optical scatterometry is a method to measure the size and shape of periodic micro- or nanostructures on surfaces. For this purpose the geometry parameters of the structures are obtained by reproducing experimental
measurement results through numerical simulations. We compare the performance of Bayesian optimization to different local minimization algorithms for this numerical optimization problem. Bayesian optimization uses Gaussian-process regression to find promising parameter values. We examine how pre-computed simulation results can be used to train the Gaussian process and to accelerate the optimization.
\end{abstract}


Copyright 2019 Society of Photo‑Optical Instrumentation Engineers (SPIE). One print or electronic copy may be made for personal use only. Systematic reproduction and distribution, duplication of any material in this publication for a fee or for commercial purposes, and modification of the contents of the publication are prohibited.

%%%%%%%%%%%%%%%%%%%%%%%%%%%%%%%%%%%%%%%%%%%%%%%%%%%%%%%%%%%%%%%%%%%%%%%%%%%%%%
% SECTIONS
%%%%%%%%%%%%%%%%%%%%%%%%%%%%%%%%%%%%%%%%%%%%%%%%%%%%%%%%%%%%%%%%%%%%%%%%%%%%%%
\section{Introduction}
\label{section_introduction}

Geometry reconstruction based on scatterometric data is a challenging numerical task. The sample structures and the measurement process are typically described using many parameters. This leads to high-dimensional optimization problems of finding shape parameters that are in agreement with the experimental data~\cite{Hammerschmidt2017,Wurm2017oe,soltwisch2017reconstructing}.

The forward-problem of computing the scattering behavior of the setup at a given point in the parameter space requires to rigorously solve Maxwell's equations. We use our finite-element method (FEM) implementation JCMsuite\cite{cite_jcmwave_jcmsuite,Pomplun2007pssb} to this aim.
The optimization process can be often solved more efficiently by providing parameter derivatives. Therefore, we compute the gradient of the solution of Maxwell’s equations by automatic differentiation~\cite{Hammerschmidt2017}.

A large number of minimization algorithms can be used to solve the inverse problem of reconstructing the shape parameters. In the considered case the objective function has only a small number of local minima such that local minimization algorithms should be very efficient. Examples for gradient-based local optimization methods are the Broyden-Fletcher-Goldfarb-Shanno
(BFGS) algorithm and its low-memory, bound-constrained extension L-BFGS-B ~\cite[and references therein]{byrd1995limited} as well as the truncated Newton method~\cite{nocedal2006large}. An example for a gradient-free method is the downhill simplex algorithm (also known as Nelder-Mead method)~\cite{doi:10.1093/comjnl/7.4.308}.

Recently, techniques from the field of machine-learning have been employed for the optimization of photonic nanostructures. For example, deep neural networks trained with thousands of simulation results have been employed as accurate models for mapping a geometry to an optical response and vice versa almost instantaneously~\cite{malkiel2018plasmonic}. 
In this work, we consider consider Gaussian processes as a method to learn the behavior of the objective function~\cite{williams1998prediction}. A popular method  that employs Gaussian processes is Bayesian optimization~\cite{shahriari2016taking}. Bayesian optimization is regularly used in machine learning applications~\cite{shahriari2016taking,Golovin:2017:GVS:3097983.3098043,aws.bayes}. In the field of nano-optics it has been, e.g., employed to optimize ring resonator-based optical filters~\cite{Rehman:16} and chiral scatterers~\cite{Gutsche18}.

Bayesian optimization derives promising parameter values by means of Bayesian inference based on \emph{all} previous function evaluations. This is in contrast to local optimization methods, which only use few of the previous data points to determine new parameters. This statistical inference can often reduce the number of required simulations~\cite{schneider2018benchmarking}.

In the context of a parameter reconstruction it is possible to compute the system response for many parameter values in advance. Provided with scatterometry data of a specific structure, the deviation between numerical and experimental response for the pre-computed parameter values can serve as training data for the Gaussian process. We investigate to which extent this training can speed up the parameter reconstruction.

\section{Scatterometric measurement configuration}
\label{section_setup}

In order to assess the optimization methods for solving the inverse problem, we use a critical dimension metrology setup studied already in previous publications \cite{Hammerschmidt2017,hammerschmidt2018solving}. 
The scatterometric measurement was executed at Physikalisch-Technische Bundesanstalt (PTB) and a
detailed investigation of the data and optical model can be found in a
recently published paper~\cite{Wurm2017oe}.

For the sake of completeness, we briefly review the experimental setup and 
the optical model: A silicon grating (1D periodic lines) with nominal pitch of 
$p_x=50$\,nm and nominal line-width of $CD=25$\,nm
was used as scattering target in a goniometric setup with an inspection 
wavelength of $\lambda=266\,$nm.  A light beam with defined polarization and 
angle of incidence (inclination angle $\theta$, rotation angle $\phi$) 
illuminates the target.
Due to the grating period and the wavelength only the zeroth diffraction order 
is present and the intensity of the reflected light in this diffraction order is recorded for S- and 
for P-polarized illumination at different inclination angles $\theta$. 
Two azimuthal rotations ($\phi=0$ and $\phi=90$\textdegree{}) are recorded.

\begin{figure}[bht]
\begin{center}
\psfrag{in}{\sffamily in}
\psfrag{out}{\sffamily out}
\psfrag{}{\sffamily }
\psfrag{hsi}{\sffamily $h$}
\psfrag{hox}{\sffamily $h_\textrm{ox}$}
\psfrag{theta}{\sffamily $\theta$}
\psfrag{phi}{\sffamily $\phi$}
\includegraphics[width=.3\textwidth]{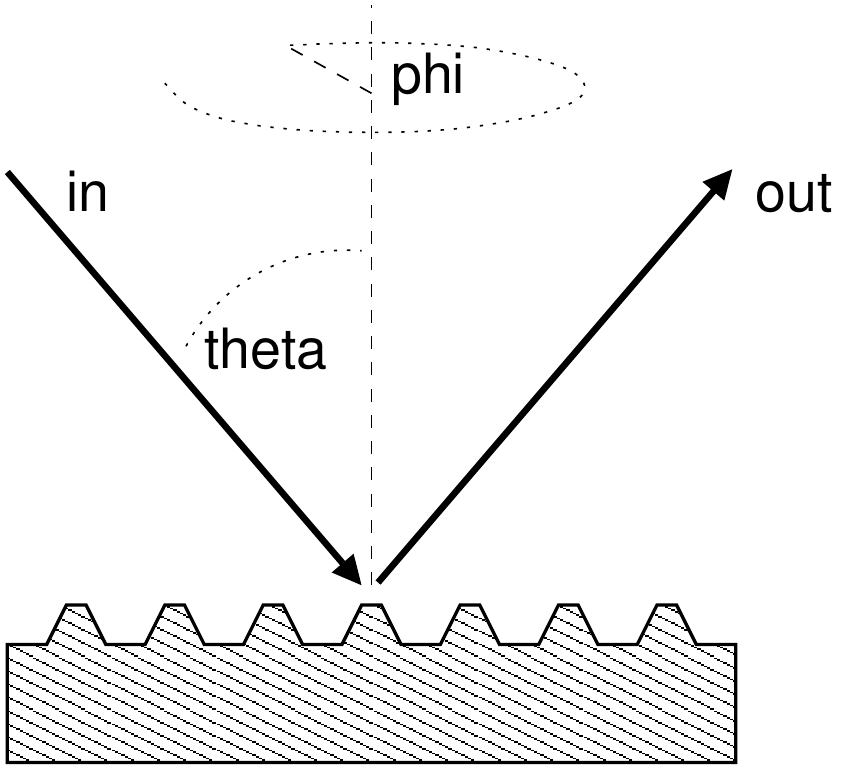} 
\includegraphics[width=.5\textwidth]{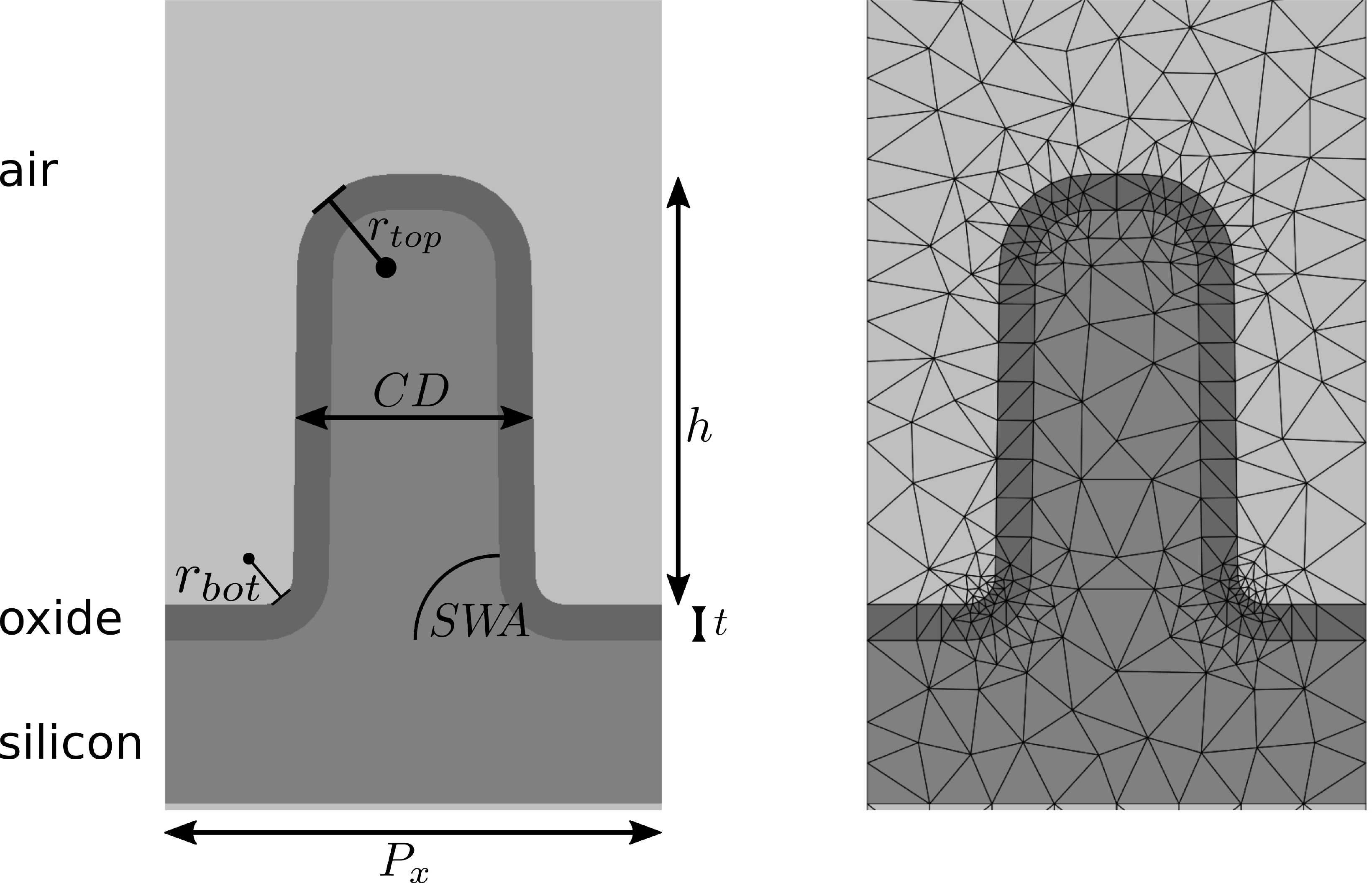}
  \caption{
{\it Left:}
Schematics of the experimental 2$\theta$ setup with incidence angle $\theta$ and 
azimuthal orientation $\phi$. 
{\it Center:}
Schematics of the model of a unit cell of the silicon line grating with free 
parameters 
line height, $h$, 
critical dimension at $h/2$, $CD$, 
oxide thickness, $t$, 
sidewall angle $SW\!\!A$, 
top and bottom corner roundings, $r_\textrm{top}$ and $r_\textrm{bot}$.
{\it Right:}
Visualization of the triangular mesh for the FEM discretization. 
}
\label{fig_schematics_scatterometry}
\end{center}
\end{figure}

\begin{figure}[htbp]
\centering
\psfrag{phi=0, P-Pol}{\sffamily \tiny $\phi=0$, P-Pol}
\psfrag{phi=90, P-Pol}{\sffamily \tiny $\phi=90$, P-Pol}
\psfrag{phi=0, S-Pol}{\sffamily \tiny $\phi=0$, S-Pol}
\psfrag{phi=90, S-Pol}{\sffamily \tiny $\phi=90$, S-Pol}
\psfrag{I_norm}{\sffamily I\textsubscript{norm}}
\psfrag{Theta [deg]}{\sffamily $\theta$ [\textdegree{}]}
\psfrag{max}{\sffamily \tiny max}
\psfrag{# unknowns}{\sffamily \# unknowns}
\includegraphics[width=.45\textwidth]{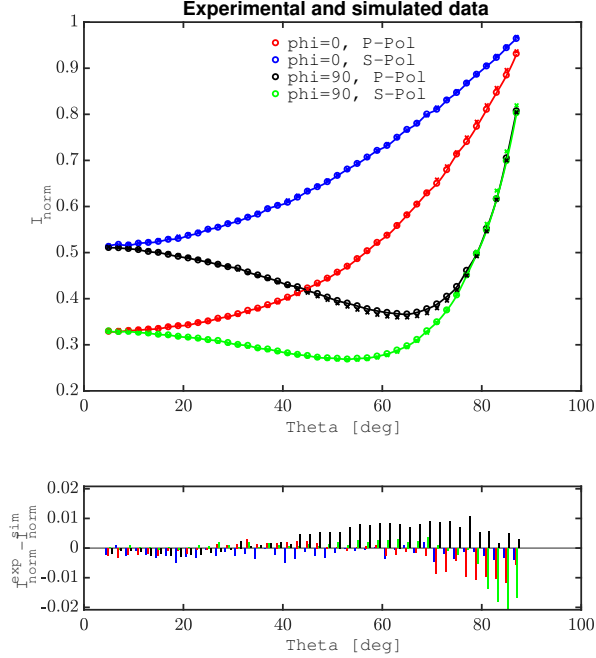}
  \caption{
Experimental data  (circles and connecting lines) and simulated data for the 
$\mathbf{x}_\textrm{MAP}$ configuration  (crosses). 
% (cf. Figure \ref{fig:grids} for the line shape).
We observe a very good quantitative alignment of the data.
The plot at the bottom
shows the difference between measured and simulated signals. Largest deviations are observed for the S-polarization,
$\phi = 90$\textdegree{} and large inclination angles $\theta$.}
\label{fig:measurement_vs_sim}
\end{figure}

A schematic of the measurement is shown in 
Figure~\ref{fig_schematics_scatterometry} (left).
The measured data set used in this study is plotted in 
Figure~\ref{fig:measurement_vs_sim} (circles). See Ref.~\cite{Hammerschmidt2017} for further explanations.

%%%%%%%%%%%%%%%%%%%%%%%%%%%%%%%%%%%%%%%%%%%%%%%%%%%%%%%%%%%%%%%%%%%%%%%%%%%%%%

\paragraph{Optical model.}
The aim of the optical model is to describe the measurements as a 
function of the micro\-structure's parameters $\mathbf{x}\in\mathbb{R}^n$. 
The shape of the silicon line is parameterized with $n=6$ free parameters: the line height  $h$, the line width (critical 
dimension) at $h/2$, $CD$, the oxide layer thickness,  $t$, the 
sidewall angle, $SW\!\!A$, and the top and bottom corner roundings, $r_\textrm{top}$ 
and $r_\textrm{bot}$. Their definitions can be found in 
Figure~\ref{fig_schematics_scatterometry} (center).  

In the reconstruction we allow for parameter values within large intervals describing a wide range of line shapes. To avoid non-physical self-intersections, we demand these to stay in an admissible bounded region. The admissible region can in general be defined as $A = \{ 
\mathbf{x}\in\mathbb{R}^m \mid g_i(\mathbf{x})\ge 0, i=1,\dots,r\}$ with smooth scalar valued 
functions $g_i$. For example, we demand that the corner rounding radius at the top of the line is smaller than half the width at the 
top. The admissible region is included into the objective function by means of a prior density as
\begin{equation}\label{eq:prior}
 \pi(\mathbf{x}) \sim \exp\left( \sum_{i=1}^r \log g_i(\mathbf{x}) \right).
\end{equation}

The scattering of monochromatic light off the nanoscopic line grating is 
described by the linear Maxwell equations in frequency domain.
These lead to a single second order partial differential equation
\begin{equation}\label{eq:mw}
 \curl \mathbf{\mu}^{-1} \curl \mathbf{E} - \omega^2 \mathbf{\varepsilon} \mathbf{E} = 0 ,
\end{equation}
where $\mathbf{\varepsilon}$ and $\mathbf{\mu}$ are the permittivity and permeability tensors, and 
$\omega$ is the time-harmonic frequency.
We employ the finite-element (FEM) electromagnetic field solver 
JCMsuite~\cite{cite_jcmwave_jcmsuite,Pomplun2007pssb}, which has been successfully used in scatterometric investigations ranging from the optical~\cite{potzick2008international} to the EUV and X-ray 
regimes~\cite{Scholze2007a,Soltwisch2016prb}
on 2D (e.g., line masks) and 3D (e.g., FinFETs, contact holes) scattering 
targets.

\paragraph{Inverse problem.} 
To reconstruct likely parameters, we use the same objective function as in \cite{Hammerschmidt2017}. 
That is, we minimize the conditional probability $\pi(\mathbf{x}|\mathbf{y}_M)$ of the parameter vector $\mathbf{x}$ given the measurement vector $\mathbf{y}_M$, also known as posterior probability. 
The posterior is given by Bayes' theorem as the product of likelihood $\pi(\mathbf{y}_M|\mathbf{x})$ and prior probability $\pi(\mathbf{x})$. 
We assume independently normally distributed measurement errors with zero mean and diagonal 
covariance $\mathbf{\Gamma}_l$:
\begin{equation}\label{eq:likelihood}
 \pi(\mathbf{y}_M|\mathbf{x}) \sim \exp\left(-\frac{1}{2} (\mathbf{y}_M-\mathbf{y}(\mathbf{x}))^T \mathbf{\Gamma}_l^{-1} (\mathbf{y}_M-\mathbf{y}(\mathbf{x}))\right).
\end{equation}
The posterior density 
\begin{equation}\label{eq:posterior}
 \pi(\mathbf{x}|\mathbf{y}_M) = \pi(\mathbf{y}_M|\mathbf{x})\pi(\mathbf{x})
\end{equation}
is maximized in order to find the most probable parameter values
\begin{equation}
\mathbf{x}_{\rm MAP} = \mathop\text{arg max}_{\mathbf{x}\in X} \pi(\mathbf{x}|\mathbf{y}_M).
\end{equation}
Equivalently, $\mathbf{x}_{\rm MAP}$ can be found by minimizing the logarithm of 
$\pi(\mathbf{x}|\mathbf{y}_M)$,
\begin{equation}\label{eq:map}
F(\mathbf{x}) = 
\frac{1}{2}(\mathbf{y}_M-\mathbf{y}(\mathbf{x}))^T \mathbf{\Gamma}_l^{-1}(\mathbf{y}_M-\mathbf{y}(\mathbf{x})) - \sum_{i=1}^r \log g_i(\mathbf{x}).
\end{equation}

As the mapping $\mathbf{y}(\mathbf{x})$ from model parameters to scatterometry measurements is nonlinear, the posterior density $\pi(\mathbf{x}|\mathbf{y}_M)$ is non-normal and can exhibit multiple local maxima. In the previous numerical study only two local maxima were found such that most of the runs of a gradient descent method converged efficiently to the global maximum $\mathbf{x}_\textrm{MAP}$ located at $CD = 25.38$~\,nm, $h=48.08$\,nm, $SW\!\!A=86.98$\textdegree{}, $t=4.94$\,nm, $r_\textrm{top}=10.37$\,nm, and $r_\textrm{bot}= 4.79$\,nm.
The local uncertainties were quantified in terms of the covariance matrix
\begin{equation}
\mathbf{\Gamma}_p = F''(\mathbf{x}_\textrm{MAP})^{-1}
\end{equation}
yielding the standard deviations $\sigma_{CD} = 
0.395$~\,nm, $\sigma_{h}=2.484$\,nm, $\sigma_{SW\!\!A}=0.999$\textdegree{}, $\sigma_{t}=0.162$\,nm, $\sigma_{r_\textrm{top}}=4.289$\,nm and $\sigma_{r_\textrm{bot}}= 3.217$\,nm~\cite{Hammerschmidt2017}.

In Figure~\ref{fig:measurement_vs_sim} the experimental and simulated intensities are shown as function
of inclination angle $\theta$. The four different angular spectra refer to the different polarizations and azimuthal orientations of the illumination. We observe an almost perfect alignment of the simulated data for $\mathbf{x} = \mathbf{x}_\textrm{MAP}$ and the PTB measurements.

\section{Bayesian optimization method}
\label{section_bayes}
The goal of every optimization algorithm is to identify the minimum (or maximum) 
of an unknown objective function $f$ in a certain design space $\mathcal{X}\subset \bbr^d$,

\begin{equation}
\mathbf{x}_{min} = \underset{\mathbf{x} \in \mathcal{X}}{\argmin} f(\mathbf{x}).
\end{equation}

The basic idea of Bayesian optimization is to treat the unknown objective as a random function, i.e. a stochastic model on a continuous domain $\mathcal{X}$. Based on the previous observation of the objective the algorithm identifies parameter values where it is expected to find a smaller function value~\cite{shahriari2016taking,SchneiderSantiagoRockstuhl2017}.

\paragraph{Gaussian processes.}
Gaussian processes (GP) are frequently used as the stochastic model in Bayesian 
optimization. A stochastic process $(X_x)_{x\in \mathcal{X}}$ is a Gaussian process if for any $N$ points $\mathbf{x}_1^*,\cdots \mathbf{x}_N^* \in \mathcal X$ the probability of the objective to be equal to $\mathbf{Y}=(y_1, \cdots, y_N)$ follows a multivariate Gaussian distribution 
\begin{equation}
\label{eq:multivariate}
 P(\mathbf{Y}^*) = 
\frac{1}{(2\pi)^{N/2}|\mathbf{\Sigma}|^{1/2}}\exp\left[-\frac{1}{2}(\mathbf{Y}
-\mathbf{m})^T\mathbf{\Sigma}^{-1}(\mathbf{Y}-\mathbf{m})\right].
\end{equation}
with a mean vector $\mathbf{m}$ and a covariance matrix $\mathbf{\Sigma}$~\cite{rasmussen2004gaussian}.

The Gaussian process is defined by a covariance function (or kernel) $k : \mathcal{X}\times 
\mathcal{X} \rightarrow \mathbb R$ 
and a mean function $\mu : \mathcal{X} \rightarrow \mathbb{R}$. 
Without prior information the mean vector $\mathbf{m}$ and a covariance matrix $\mathbf{\Sigma}$ evaluate to 
$\mathbf{m} = [\mu(\mathbf{x}_1^*),\cdots,\mu(\mathbf{x}_N^*)]^T$ and 
$(\mathbf{\Sigma})_{ij} = k(\mathbf{x}_i^*, \mathbf{x}_j^*)$.
After $M$ iterations, the function values $Y=(y_1,\cdots, y_M)=(f(x_1),\cdots,f(x_M))$ are known. Using Bayesian inference, the mean vector and the covariance matrix then reads
\begin{eqnarray}
\label{eq:m}
\mathbf{m} &=& \mathbf{m}_2 + \mathbf{\Sigma}_{21}\mathbf{\Sigma}_{11}^{-1}(\mathbf{Y}_1^* - \mathbf{m}_1) \\
\label{eq:Sigma}
\mathbf{\Sigma} &=& \mathbf{\Sigma}_{22} - \mathbf{\Sigma}_{21}\mathbf{\Sigma}_{11}^{-1}\mathbf{\Sigma}_{21}^T
\end{eqnarray}
with $\mathbf{m}_1 = [\mu(\mathbf{x}_1),\cdots,\mu(\mathbf{x}_M)]^T$,
$\mathbf{m}_2 = [\mu(\mathbf{x}_1^*),\cdots,\mu(\mathbf{x}_N^*)]^T$, 
$(\mathbf{\Sigma}_{11})_{ij} = k(\mathbf{x}_i, \mathbf{x}_j)$,
$(\mathbf{\Sigma}_{22})_{ij} = k(\mathbf{x}_i^*, \mathbf{x}_j^*)$,
$(\mathbf{\Sigma}_{21})_{ij} = k(\mathbf{x}_i^*, \mathbf{x}_j)$, and
$(\mathbf{\Sigma}_{12})_{ij} = k(\mathbf{x}_i, \mathbf{x}_j^*)$.

Gaussian processes can be easily extended to incorporate not only data on the objective function but also of its derivatives~\cite{solak2003derivative}. For example, a gradient observation $\nabla f(\mathbf{x}_k)$ 
leads to additional entries in the covariance matrix formed by derivatives of the covariance function $\nabla_{\mathbf{x}_k}k(\mathbf{x}_i, \mathbf{x}_k)$ 
and $\nabla_{\mathbf{x}_i}\nabla_{\mathbf{x}_k} k(\mathbf{x}_i, \mathbf{x}_k)$. 

From Eqs.~(\ref{eq:multivariate}--\ref{eq:Sigma}) follows that at any position $\mathbf{x}^* \in \mathcal{X}$ the unknown function value is normally distributed, 
i.e. $f(\mathbf{x}^*)\sim \mathcal{N}(\overline{y},\sigma^2)$ with
\begin{eqnarray}
\overline{y}(\mathbf{x}^*) &=& \mu(\mathbf{x}^*) + \sum_{i j} k(\mathbf{x}^*,\mathbf{x}_i) (\Sigma_{11}^{-1})_{i j} [f(\mathbf{x}_j) - \mu(\mathbf{x}_j)]\\
\sigma(\mathbf{x}^*)^2 &=& k(\mathbf{x}^*,\mathbf{x}^*) - \sum_{i j} k(\mathbf{x}^*,\mathbf{x}_i) (\Sigma_{11}^{-1})_{i j} k(\mathbf{x}_j,\mathbf{x}^*) .
\end{eqnarray}

Based on the normal distribution the next parameters values are chosen to maximize a specific acquisition function. A popular choice is the \emph{expected improvement}, i.e. the expectation value of the improvement $\max(0,y_{\rm min} - f(\mathbf{x}^*))$ with respect to the currently known minimal function value $y_{\rm min}$ 
\begin{equation}
\label{eq:EI}
\begin{split}
\alpha_{\rm EI}(\mathbf{x}^*,y_{\rm min}) &= \mathbb E[ \max(0,y_{\rm min} - f(\mathbf{x}^*)) ] \\
&= \frac{1}{2} \left[
    1+ \rm{erf}\left(
       \frac
       	{y_{\rm min} - \overline{y}(\mathbf{x}^*)}
       	{\sqrt{2}\sigma(\mathbf{x}^*)}
    \right)
 \right]
 (y_{\rm min} - \overline{y}(\mathbf{x}^*)) + 
 \frac{\sigma(\mathbf{x}^*)}{\sqrt{2\pi}} \exp\left(
	-\frac
       	{(y_{\rm min} - \overline{y}(\mathbf{x}^*))^2}
       	{2\sigma(\mathbf{x}^*)^2} 
 \right) .
 \end{split}
\end{equation} 
Other common acquisition functions are the \emph{probability of improvement} or the \emph{lower confidence bound}~\cite{shahriari2016taking}.

\paragraph{Hyperparameter choice.}

Using different covariance functions $k(\mathbf{x},\mathbf{x}')$, Gaussian processes allow to approximate a large class of random functions. 
A popular choice is the Mat\'{e}rn $5/2$ covariance function 
\begin{equation}
k(\mathbf{x},\mathbf{x}') = s^2 \left(1 + \sqrt{5} r(\mathbf{x},\mathbf{x}') + \frac{5}{3}r(\mathbf{x},\mathbf{x}')^2\right)\exp\left(-\sqrt{5} r(\mathbf{x},\mathbf{x}')\right)\;\; \text{with}\;\; r(\mathbf{x},\mathbf{x}')^2 = \sum_i \frac{(\mathbf{x}_i - \mathbf{x}'_i)^2}{l_i^2} .
\end{equation}
Moreover, we choose a constant mean function $\mu(\mathbf{x}) = \mu_0$. The values of the hyperparameters $\mathbf{w}= (\mu0, s^2,$ $l_1^2, \cdots,l_d^2)$ are essential for the performance of the optimization procedure.
The idea is to maximize the likelihood $P(\mathbf{Y}) = P_\mathbf{w}(\mathbf{Y})$ of all known objective function values with respect to the hyperparameters. This hyper parameter optimization is computationally expensive and is only performed if the derivatives of $P_\mathbf{w}(\mathbf{Y})$ with respect to the length scales exceed a certain threshold~\cite{SantiagoSchneiderRockstuhl2018}.

\paragraph{Learning from offline calculations.}

In a typical parameter reconstruction setup many specimens with the same type of geometry have to be probed. Therefore, we examine whether the reconstruction process can be accelerated by pre-comuting the optical model $\mathbf{y}(\mathbf{x})$ for many parameter values $\mathbf{x} \in X$. That is, provided with a measurement result $\mathbf{y}_M$ the parameters $X$ and the corresponding posterior probabilities $P_0 = \{\pi(\mathbf{x}|\mathbf{y}_M) \;|\; \mathbf{x}\in X\}$ and the parameter derivatives $P_i = \{\partial\pi(\mathbf{x}|\mathbf{y}_M)/\partial \mathbf{x}_i \;|\; \mathbf{x}\in X\}$ for $i=1,\cdots,6$ are used to initialize the Gaussian process underlying Bayesian optimization. We draw $X$ from a pseudo-random Sobol sequence and use only parameters that meet the constraints, i.e. $X\subset A$~\cite{sobol1967distribution,sobol_py}.

\section{Results}
\label{section_results}

In order to assess the performance of the different optimization approaches we have conducted six independent optimization runs of 150 iterations with different initial conditions for each method. That is, the Downhill-Simplex method, L-BFGS-B, Newton Conjugate-Gradient, and Bayesian optimization were started from six different random initial points. Whenever a local minimization method converged to a local minimum, it was restarted at a different position. Moreover, six independent sets of points $(X^{(i)})_{i=1,\cdots,6}$ of 100 training samples and corresponding posterior probabilities and derivatives $(P_0^{(i)}, \cdots, P_6^{(i)})_{i=1,\cdots,6}$  were prepared. The sets were used to initialize the Gaussian processes ("Bayesian optimization + training"). Apart from the downhill simplex methods, all methods make use of derivative information to determine the next sampling point.

\begin{figure}[bht]
\begin{center}
\includegraphics[width=.45\textwidth]{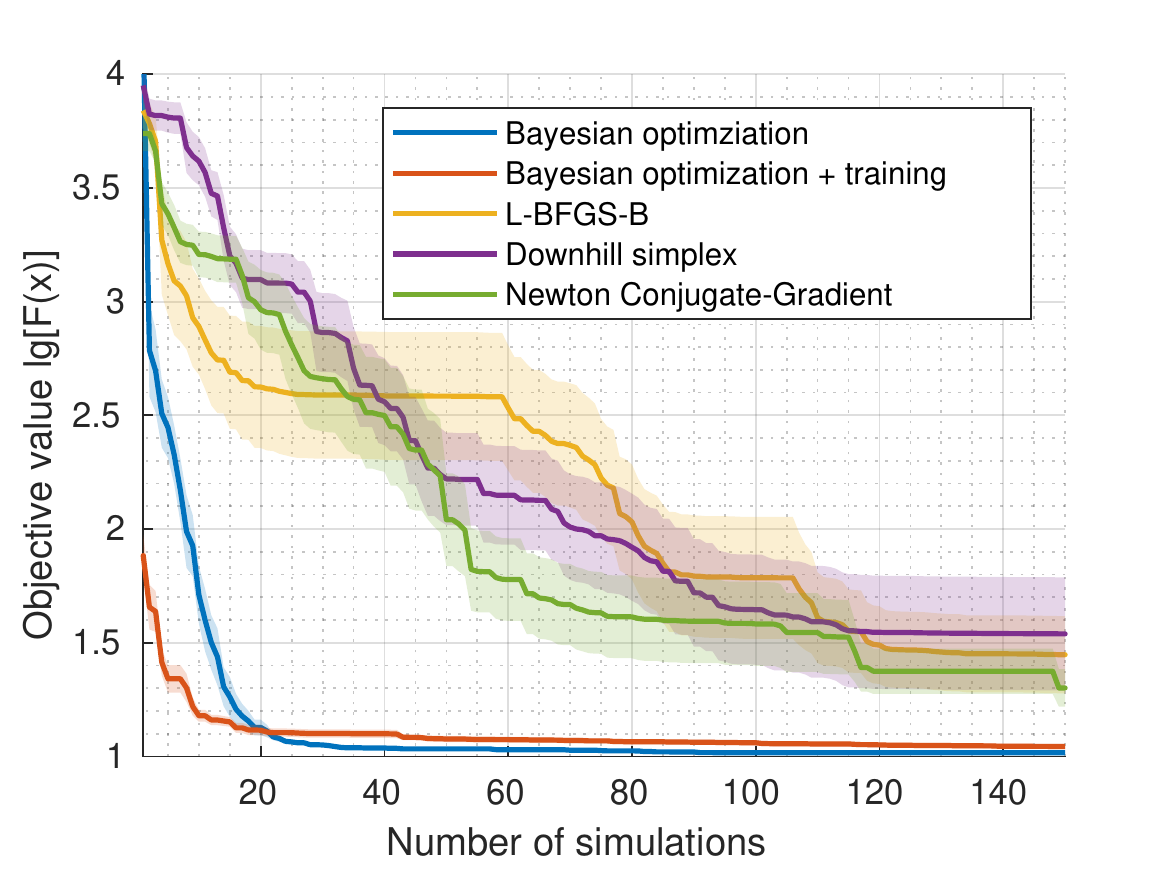}
\includegraphics[width=.45\textwidth]{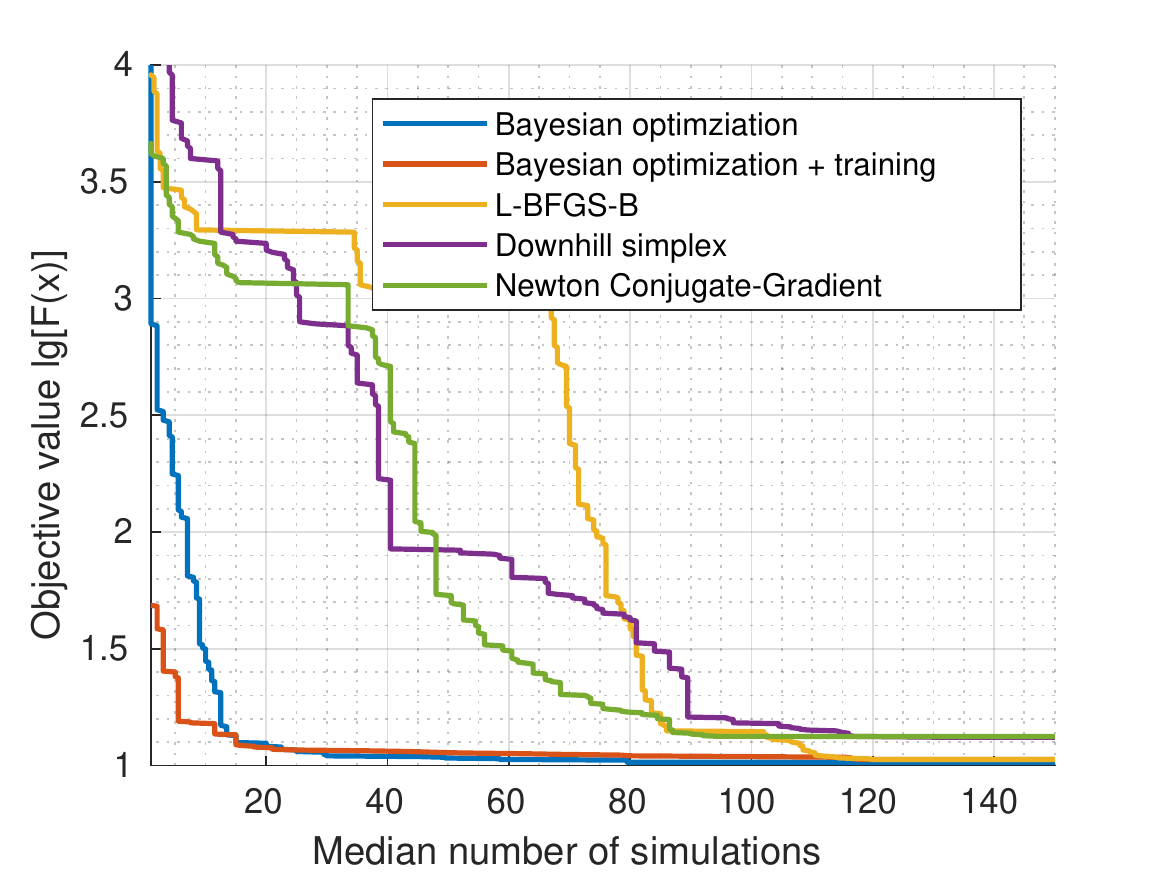}
  \caption{
{\it Left:} Average objective value $\lg[F(\mathbf{x})]$ as a function of the number of simulations for six independent optimization runs for each method. The shaded areas indicate the standard deviation between the six optimization runs.
{\it Right:} Median number of simulations needed to obtain an objective value smaller or equal to $\lg[F(\mathbf{x})]$ shown on the $y$-axis.
}
\label{fig_error}
\end{center}
\end{figure}

Since the error function $F(\mathbf{x})$ [see Eq.~\eqref{eq:map}] varies between several orders of magnitude, we minimize its decadic logarithm $\lg[F(\mathbf{x})]$. Figure~\ref{fig_error} compares the performance of the optimization approaches for minimizing $\lg[F(\mathbf{x})]$. On the left, the average objective value is shown as a function of the number of simulations for each of the different optimization methods. Because some optimization runs can fail by being trapped into a local minimum, the average objective value is not always meaningful. Therefore, the right plot shows also the median number of simulations needed to obtain objective values below a certain threshold. The median is less sensitive against failed runs, as long as at least four of the 6 runs are successful. Bayesian optimization needs on average significantly less simulations to find the minimum than the local optimization methods.
Surprisingly, the performance of the derivative-free downhill-simplex method is similar  to the gradient descent methods in minimizing $\lg[F(\mathbf{x})]$. Only close to the global minimum L-BFGS-B converges better to objective values below 1.1.
Clearly, the 100 training samples lead to a significant speed-up of Bayesian optimization. Only after about 20 iterations the non-trained approach converges faster to the global minimum. We attribute this to the following behavior: Both Bayesian optimization methods converge into a region close to the global minimum. Because of small parameter derivatives, the expected improvement in this region becomes very small. Provided with many training samples the second approach now identifies other regions in the parameter space where some improvement can be expected and tends to explore these regions. If no training samples are available, the same happens at a later stage.

\begin{figure}[bht]
\begin{center}
\includegraphics[width=.45\textwidth]{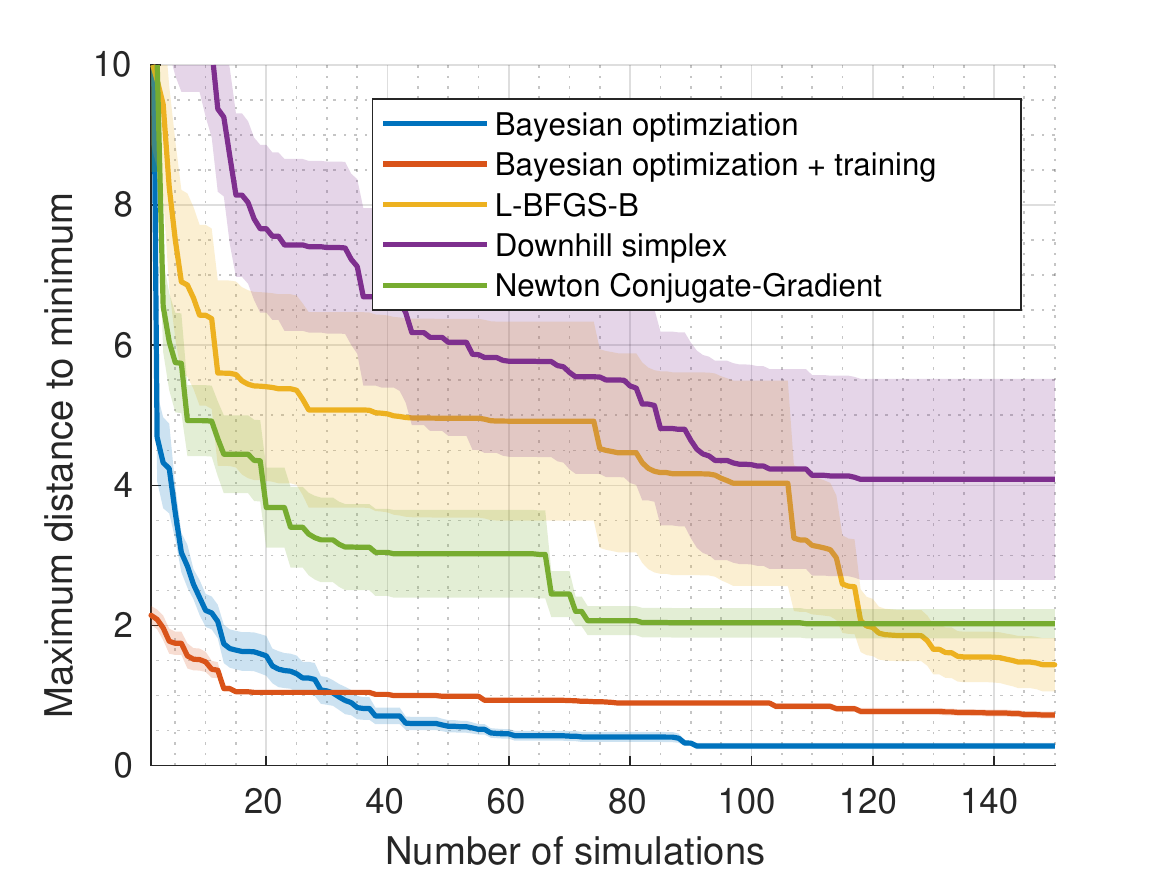} 
\includegraphics[width=.45\textwidth]{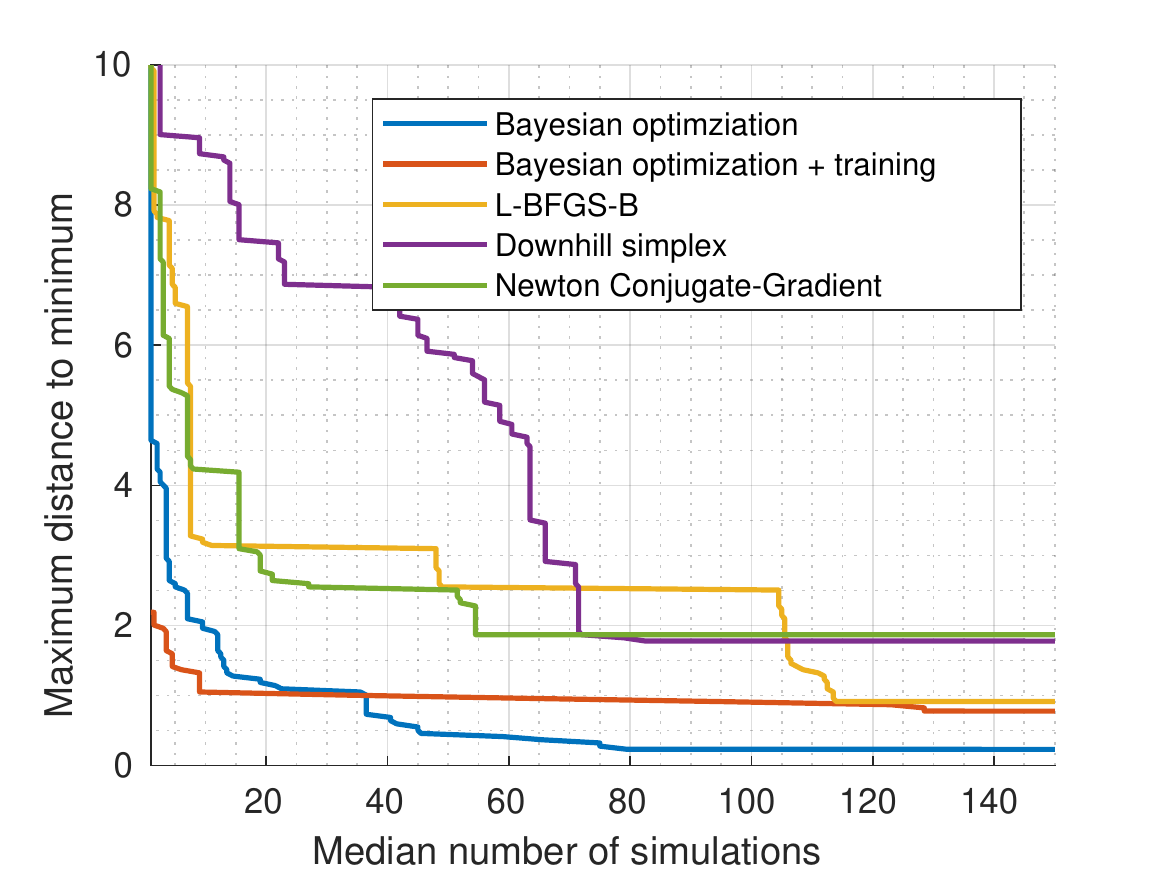}
  \caption{
{\it Left:} Average maximum distance to global minimum as a function of the number of simulations for six independent optimization runs for each method. The shaded areas indicate the standard deviation between the six optimization runs.
{\it Right:} Median number of simulations needed to converge to a region with a specific maximum distance to the global minimum shown on the $y$-axis.
}
\label{fig_distance_to_minimum}
\end{center}
\end{figure}

Figure~\ref{fig_distance_to_minimum} shows the maximum distance to the global minimum as a function of the number of simulations. The distance is measured in units of the measurement uncertainty for all six geometry parameters, i.e.,
\begin{equation}
d = \max_{i=1,\cdots,6} \frac{\left|(\mathbf{x} - \mathbf{x}_{\rm MAP})_i \right|}{\mathbf{\sigma}_{i}}
\end{equation}
Bayesian optimization with training converges after a median number of 9 iterations to a region within the measurement uncertainty ($d=1$), i.e. where the derivatives of $F(\mathbf{x})$ become small. Without training, Bayesian optimization needs more than 25 iterations to converge to the same accuracy level. From the local minimization methods, only L-BFGS-B converges to the measurement uncertainty within 150 simulations.

%%%%%%%%%%%%%%%%%%%%%%%%%%%%%%%%%%%%%%%%%%%%%%%%%%%%%%%%%%%%%%%%%%%%%%%%%%%%%%

%%%%%%%%%%%%%%%%%%%%%%%%%%%%%%%%%%%%%%%%%%%%%%%%%%%%%%%%%%%%%%%%%%%%%%%%%%%%%%
% CONCLUSION ETC
%%%%%%%%%%%%%%%%%%%%%%%%%%%%%%%%%%%%%%%%%%%%%%%%%%%%%%%%%%%%%%%%%%%%%%%%%%%%%%
\section{Conclusion}
\label{section_conclusion}

We have compared the performance of Bayesian optimization to different local minimization algorithms for a specific example of a geometry parameter reconstruction based on scatterometry data. We find that Bayesian optimization finds the most probable parameters with a significantly smaller number of simulation results. A training with pre-computed simulations can further speed up the reconstruction up to the point where parameter values within the measurement uncertainty are identified. Thereafter, the convergence speed decreases since the expected improvement around other training samples is larger. Other acquisition functions such as the probability of improvement could prevent the exploration of other parts of the parameter space. In order to improve the convergence speed it might be beneficial to change to a different Bayesian optimization strategy or to a gradient descent method when the expected improvement becomes small.  

\section*{Acknowledgments}
This work is partially funded through the project 17FUN01 "BeCOMe" within the Programme EMPIR. The EMPIR initiative is co-founded by the European Union's Horizon 2020 research and innovation program and the EMPIR Participating Countries.
Further, we acknowledge support by the Central Innovation Programme for SMEs of the German Federal Ministry for Economic Affairs and Energy on the basis of a decision by the German Bundestag (ZF4450901).

\end{document}